\documentclass[twocolumn,english]{revtex4-2}
\usepackage[T1]{fontenc}
\usepackage[latin9]{luainputenc}
\usepackage{babel}
\usepackage{amstext}
\usepackage{amssymb}
\usepackage{graphicx}
\usepackage[unicode=true,pdfusetitle,
 bookmarks=true,bookmarksnumbered=false,bookmarksopen=false,
 breaklinks=false,pdfborder={0 0 1},backref=false,colorlinks=false]
 {hyperref}
\begin{document}
\title{Meta-stable states in the Ising model with Glauber-Kawasaki competing
dynamics }
\author{R. A. Dumer }
\email{rafaeldumer@fisica.ufmt.br}

\affiliation{Instituto de Física - Universidade Federal de Mato Grosso, 78060-900,
Cuiabá, Mato Grosso, Brazil.}
\author{M. Godoy}
\email{mgodoy@fisica.ufmt.br}

\affiliation{Instituto de Física - Universidade Federal de Mato Grosso, 78060-900,
Cuiabá, Mato Grosso, Brazil.}
\begin{abstract}
Meta-stable states are identified in the Ising model with competition
between the Glauber and Kawasaki dynamics. The model of interaction
between magnetic moments was implemented on a network where the degree
distribution follows a power-law of the form, $P(k)\sim k^{-\alpha}$.
The evolution towards the stationary state occurred through the competition
between two dynamics, driving the system out of equilibrium. In this
competition, with probability $q$, the system was simulated in contact
with a heat bath at temperature $T$ by the Glauber dynamics, while
with probability $1-q$, the system experienced an external energy
influx governed by the Kawasaki dynamics. The phase diagrams of $T$
versus $q$ were obtained, which are dependent on the initial state
of the system, and exhibit first- and second-order phase transitions.
In all diagrams, for intermediate values of $T$, the phenomenon of
self-organization between the ordered phases was observed. In the
regions of second-order phase transitions, we have verified the universality
class of the system through the critical exponents of the order parameter
$\beta$, susceptibility $\gamma$, and correlation length $\nu$.
Furthermore, in the regions of first-order phase transitions, we have
demonstrated the instability due to transitions between the ordered
phases through hysteresis-like curves of the order parameter, in addition
to the existence of absorbing states. We also estimated the value
of the tricritical points when the discontinuity in the order parameter
in the phase transitions was no longer observed.
\end{abstract}
\maketitle

\section{Introduction}

One of the main points of interest when dealing with non-equilibrium
systems is the possibility of finding phase transitions with characteristics
of reversible systems, even though in this non-equilibrium thermodynamic
regime, we lack a unifying framework like a Gibbs equilibrium statistical
mechanics \citep{1}. We can handle non-equilibrium systems when the
evolution process toward the steady state involves competition between
two dynamics \citep{2,3,4,5}. This is because individually these
dynamics satisfy detailed balance, but if both have a non-zero probability
of acting on the system, the principle of microscopic reversibility
is not always respected, and the system is forced out of equilibrium.
Two dynamics that are commonly employed in competition are the Glauber
dynamics with the single-spin flip process \citep{6} and the Kawasaki
dynamics with the two-spin exchange process \citep{7}.

Given its simplicity and usefulness in studying phase transitions,
the Ising model is also widely employed in investigating non-equilibrium
systems with competitive dynamics. In such cases, considering a ferromagnetic
coupling between spins, with probability $q$, the system is in contact
with a thermal reservoir at temperature $T$ and it relaxes to the
steady state of lower energy through the Glauber dynamics. On the
other hand, with probability $1-q$, the system is subject to an external
energy influx, and evolves to the state of higher energy through the
Kawasaki dynamics. In a regular square lattice \citep{8}, the Ising
model subject to these competing dynamics self-organizes into the
ordered phases, the ferromagnetic phase ($F$) and the antiferromagnetic
phase ($AF$). In this self-organization, at low values of $q$, the
$AF$ phase is found, corresponding to the higher energy state of
the system. In contrast, when $q$ increases, a phase transition occurs
to the paramagnetic phase ($P$). Further increasing $q$ leads to
another transition to an ordered phase, the $F$ phase, corresponding
to the lower energy state of the system. All these transition lines
are of second-order.

Beyond regular networks, complex networks despite not having much
evidence to describe crystals, are of great interest because they
describe a range of structures found in society. Examples of these
are the small-world networks \citep{9,10}, which encompass the property
discovered by Milgram \citep{11}, wherein any person in the world
can have contact with another, requiring a remarkably smaller number
of intermediaries compared to the size of the network. Another example
of complex networks present in society are those that follow a power-law
degree distribution, $p(k)\sim k^{-\alpha}$ \citep{12}. In this
case, $p(k)$ is the probability of any point in the network having
$k$ other points connected to it, and $\alpha$ is the exponent that
depends on the object of study. Networks of this kind are notable
due to advancements in data processing techniques and equipment. It
has been observed that networks such as the World Wide Web, the Internet,
citation networks, networks of actors who have appeared in the same
film, networks of protein interactions, among many others \citep{13},
despite having distinct formation origins, self-organize so that the
degree distribution takes the form of a power law.

Due to the importance of complex networks, they have been implemented
in physical models to investigate their influence on phase transitions
\citep{14,15,16,17}. These models also include the non-equilibrium
Ising model through competitive dynamics. The phenomenon of self-organization
is observed with competing dynamics of one- and two-spin flips on
small-world networks \citep{18} and networks with power-law degree
distribution \citep{19}. This involves transitions from the $AF$
to $P$ phases and from the $P$ to $F$ phases, varying the competition
parameter $q$. In both cases, only second-order phase transitions
are found, and the universality class obtained through critical exponents,
in both networks, with and without competitive dynamics, belongs to
the mean-field regime \citep{20,21}. The results are also available
regarding the Ising model on a 2D small-world network and with competition
between the Glauber and Kawasaki dynamics \citep{22}. In this case,
the self-organization was also observed, but in the region of phase
diagrams $(T\times q)$, where competition between the $F$ and $AF$
ordered phases are present, i.e., low values of $q$ and $T$, first-order
phase transitions are found, in addition to second-order phase transitions
for low values of $q$ with high values of $T$, and high values of
$q$ with low values of $T$.

In the present work, we investigated the Ising model on a network
with a power-law degree distribution and with competition between
the Glauber and Kawasaki dynamics. In this configuration, each point
of the network represents a spin variable that can take values of
$\sigma\pm1$, with a probability $p(k)\sim k^{-1}$ of interacting
with $k$ other spins randomly distributed in the network. For the
evolution towards the steady state, with probability $q$, the system
is in contact with a thermal reservoir at temperature $T$ and it
relaxes to the lowest energy state through Glauber dynamics. Meanwhile,
with a probability $1-q$ , there is an external energy flux into
the system, governed by Kawasaki dynamics, favoring the higher energy
state. Thus, we aim to fill a gap in the study of non-equilibrium
systems due to competitive dynamics in complex networks. We have found
first-order phase transition lines in the Ising model on a complex
network, as well as rich phase diagrams dependent on the initial state
of the system, given the diffusive dynamic involved, with $AF$, $F$,
and $P$ phases, and tricritical points separating first-order phase
transitions from second-order ones.

This article is organized as follows: In Section \ref{sec:Model},
we present the network, the dynamics involved in the system, and how
they drive the evolution of the Ising system. In Section \ref{sec:Monte-Carlo-simulations},
we provide details about the Monte Carlo method, the thermodynamic
quantities of interest, and the scaling relations for each of them.
The phase diagrams and a detailed description of both first- and second-order
phase transitions present in these diagrams are discussed in Section
\ref{sec:Results}. Finally, in Section \ref{sec:Conclusions}, we
present the conclusions drawn from the study.

\section{Model\label{sec:Model}}

Here, we have utilized a network divided into two sublattices. The
sites from one sublattice can only randomly connect to spins of the
other sublattice, and the degree of the sites follows a power-law
distribution of the form
\begin{equation}
p(k)=\frac{k^{-\alpha}}{\sum_{k=k_{0}}^{k_{m}}k^{-\alpha}}.\label{eq:1}
\end{equation}
With $k_{0}=4$ being the minimum degree of the sites, $k_{m}=10$
the maximum degree present on the network, and a fixed value of $\alpha=1$,
we impose limitations on the degrees of the network, thereby disrupting
the scale-free network property typically found in real networks exhibiting
growth and preferential connections \citep{12}. This is done to ensure
finite critical points and mean-field universality class, as our focus
in this study is on the effects of reactive-diffusive competing dynamics
on the well-defined network implemented in the Ising model. Further
details on the network construction and the effect of the exponent
$\alpha$ at the criticality of the system can be found in two previous
works \citep{19,21}.

In the Ising model, the interaction energy between the spins is defined
by the Hamiltonian in the form

\begin{equation}
\mathcal{H}=-\sum_{\left\langle i,j\right\rangle }J_{ij}\sigma_{i}\sigma_{j}\label{eq:2}
\end{equation}
where $\sigma_{i}=\pm1$, the sum is over all pair of spins, and we
use $J_{ij}=1$, meaning ferromagnetic interaction if sites $i$ and
$j$ interact between the sublattices, and zero otherwise.

In the non-equilibrium system studied here, let us denote $p(\{\sigma\},t)$
as the probability of finding the system in the state $\{\sigma\}=\{\sigma_{1},...,\sigma_{i},...,\sigma_{j},...\sigma_{N}\}$
at time $t$. The equation governing the evolution of the probability
of states over time is given by the master equation

\begin{equation}
\frac{d}{dt}p(\{\sigma\},t)=qG+(1-q)K,\label{eq:3}
\end{equation}
where $qG$ represents the one-spin flip process, associated with
the Glauber dynamics, which relaxes the spins in contact with a heat
bath at temperature $T$, favoring the lowest energy state of the
system, and it have probability $q$ to occur. On the other hand,
$(1-q)K$ represents the two-spin exchange process, related with the
Kawasaki dynamics, where the system is subjected to an external flux
of energy into it, increasing the energy of the system, and it have
probability $1-q$ to occur. $G$ and $K$ are described as follows:

\begin{equation}
\begin{array}{ccc}
G= & \sum_{i,\{\sigma'\}}\left[W(\sigma_{i}\to\sigma_{i}')p(\{\sigma\},t)+\right.\\
 & \left.-W(\sigma_{i}'\to\sigma_{i})p(\{\sigma'\},t)\right] & ,
\end{array}\label{eq:4}
\end{equation}

\begin{equation}
\begin{array}{ccc}
K= & \sum_{i,j,\{\sigma'\}}\left[W(\sigma_{i}\sigma_{j}\to\sigma_{j}\sigma_{i})p(\{\sigma\},t)+\right.\\
 & \left.-W(\sigma_{j}\sigma_{i}\to\sigma_{i}\sigma_{j})p(\{\sigma'\},t)\right] & ,
\end{array}\label{eq:5}
\end{equation}
where $\{\sigma'\}$ is the new the spin configuration, $W(\sigma_{i}\to\sigma_{i}')$
is the transition rate between the states on the one-spin flip process,
and $W(\sigma_{i}\sigma_{j}\to\sigma_{j}\sigma_{i})$ the transition
rate between the states in the two-spin exchange process.

\section{Monte Carlo simulations\label{sec:Monte-Carlo-simulations}}

In our Monte Carlo simulations, we have considered two possible initial
states for the system: the ordered state, where all the spins are
the same state, and the disordered state, where the spin states are
randomly chosen. Starting from the initial state, a new spin configuration
is generated following the Markov process: for a given temperature
$T$, competition probability $q$, and network size $N=L\times L$,
we randomly select a spin $\sigma_{i}$ in the network and generate
a random number $r$, uniform distributed between zero and one. If
$r\le q$, we choose the one-spin flip process, in which the flipping
probability is given by the Metropolis prescription:

\begin{equation}
W(\sigma_{i}\to\sigma_{i}^{\prime})=\left\{ \begin{array}{cccc}
e^{\left(-\Delta E_{i}/k_{B}T\right)} & \textrm{if} & \Delta E_{i}>0\\
1 & \textrm{if} & \Delta E_{i}\le0 & ,
\end{array}\right.\label{eq:6}
\end{equation}
where $\Delta E_{i}$ is the change in energy, based in Eq. (\ref{eq:2}),
after flipping the spin $\sigma_{i}$, $k_{B}$ is the Boltzmann constant,
and $T$ the temperature of the system. In summary, a new state is
accepted if $\Delta E_{i}\le0$. However, if $\Delta E>0$ the acceptance
is determined by the probability $\exp\left(-\Delta E_{i}/k_{B}T\right)$,
and it is accepted only if a randomly chosen number $r_{1}$uniformly
distributed between zero and one satisfies $r_{1}\le\exp\left(-\Delta E_{i}/k_{B}T\right)$.
If none of these conditions are satisfied, the state of the system
remains unchanged. Now, if $r>q$ the two-spin exchange process is
chosen, and in addition to the spin $\sigma_{i}$ we also randomly
choose one of its neighbors $\sigma_{j}$, and the state of these
two spins are exchanged according to transition rate

\begin{equation}
W(\sigma_{i}\sigma_{j}\to\sigma_{j}\sigma_{i})=\left\{ \begin{array}{c}
0\\
1
\end{array}\begin{array}{c}
\textrm{if}\\
\textrm{if}
\end{array}\begin{array}{cc}
\Delta E_{ij}\le0\\
\Delta E_{ij}>0 & ,
\end{array}\right.\label{eq:7}
\end{equation}
where $\Delta E_{ij}$ is the change in the energy after exchange
the state of the spins $\sigma_{i}$ and $\sigma_{j}$. In this process,
the new state is accepted only if the change in the energy is greater
than zero. Whereas this process aims to simulate the system under
the influence of an external energy input, thus an increase in energy
is expected.

Repeating the Markov process $n$ times, we have one Monte Carlo Step
(MCS). We allowed the system to evolve for $n=10^{4}$ MCS to reach
a stationary state, for all network sizes, $(32)^{2}\leq N\leq(256)^{2}$.
To calculate the thermal averages of the quantities of interest, we
conducted an additional $4\times10^{4}$ MCS, and the averaging across
samples was performed using $10$ independent samples for each configuration.

The measured thermodynamic quantities in our simulations are: magnetization
per spin $\textrm{m}_{\textrm{N}}^{\textrm{F}}$, staggered magnetization
per spin $\textrm{m}_{\textrm{N}}^{\textrm{AF}}$, magnetic susceptibility
$\textrm{\ensuremath{\chi}}_{\textrm{N}}$ and reduced fourth-order
Binder cumulant $\textrm{U}_{\textrm{N}}$:

\begin{equation}
\textrm{m}_{\textrm{N}}^{\textrm{F}}=\frac{1}{N}\left[\left\langle \sum_{i=1}^{N}\sigma_{i}\right\rangle \right],\label{eq:8}
\end{equation}

\begin{equation}
\textrm{m}_{\textrm{N}}^{\textrm{AF}}=\frac{1}{N}\left[\left\langle \sum_{i=1}^{N}(-1)^{(l+c)}\sigma_{i}\right\rangle \right],\label{eq:9}
\end{equation}

\begin{equation}
\textrm{\ensuremath{\chi}}_{\textrm{N}}=\frac{N}{k_{B}T}\left[\left\langle m^{2}\right\rangle -\left\langle m\right\rangle ^{2}\right],\label{eq:10}
\end{equation}

\begin{equation}
\textrm{U}_{\textrm{N}}=1-\frac{\left[\left\langle m^{4}\right\rangle \right]}{3\left[\left\langle m^{2}\right\rangle ^{2}\right]},\label{eq:11}
\end{equation}
where $\left[\ldots\right]$ representing the average over the samples,
and $\left\langle \ldots\right\rangle $ the thermal average over
the MCS in the stationary state. To facilitate the calculation of
$\textrm{m}_{\textrm{N}}^{\textrm{AF}}$, the sites on the network
are labeled as if we had a square lattice, $N=L^{2}$, in this way,
$l$ and $c$ are the row and column of the site $i$, respectively.
In Eqs. (\ref{eq:10}) and (\ref{eq:11}), $m$ can represent either
$\textrm{\ensuremath{\textrm{m}_{\textrm{N}}^{\textrm{F}}}}$ or $\textrm{\ensuremath{\textrm{m}_{\textrm{N}}^{\textrm{AF}}}}$.

Near the stationary critical point $\Gamma_{c}$, the equations (\ref{eq:8}),
(\ref{eq:9}), (\ref{eq:10}) and (\ref{eq:11}) obey the following
finite-size scaling relations \citep{23}:

\begin{equation}
\textrm{m}_{\textrm{N}}=N^{-\beta/\nu}m_{0}(N^{1/\nu}\epsilon),\label{eq:12}
\end{equation}

\begin{equation}
\textrm{\ensuremath{\chi}}_{\textrm{N}}=N^{\gamma/\nu}\chi_{0}(N^{1/\nu}\epsilon),\label{eq:13}
\end{equation}

\begin{equation}
\textrm{U}_{\textrm{N}}^{\prime}=N^{1/\nu}\frac{U_{0}^{\prime}(N^{1/\nu}\epsilon)}{\Gamma_{c}},\label{eq:14}
\end{equation}
where $\epsilon=(\Gamma-\Gamma_{c})/\Gamma_{c}$ ($\Gamma$ can be
$T$ or $q$), $\beta$, $\gamma$ and $\nu$ are the critical exponents
related the magnetization, susceptibility and length correlation,
respectively. The functions $m_{0}(N^{1/\nu}\epsilon)$, $\chi_{0}(N^{1/\nu}\epsilon)$
and $U_{0}(N^{1/\nu}\epsilon)$ are the scaling functions.

\section{Results \label{sec:Results}}

This section is divided into four subsections. The subsections \ref{subsec:Ordered-initial-state}
and \ref{subsec:Disordered-initial-state} are related to the phase
diagrams obtained for the system with two different initial states.
The subsections \ref{subsec:Second-order-phase-transitions} and \ref{subsec:First-order-phase-transitions}
concern to the different phase transitions types found in the phase
diagrams, namely, second- and first-order phase transitions, respectively.

\subsection{Ordered initial state\label{subsec:Ordered-initial-state}}
\begin{center}
\begin{figure*}
\begin{centering}
\includegraphics[viewport=20bp 0bp 385bp 300bp,clip,scale=0.6]{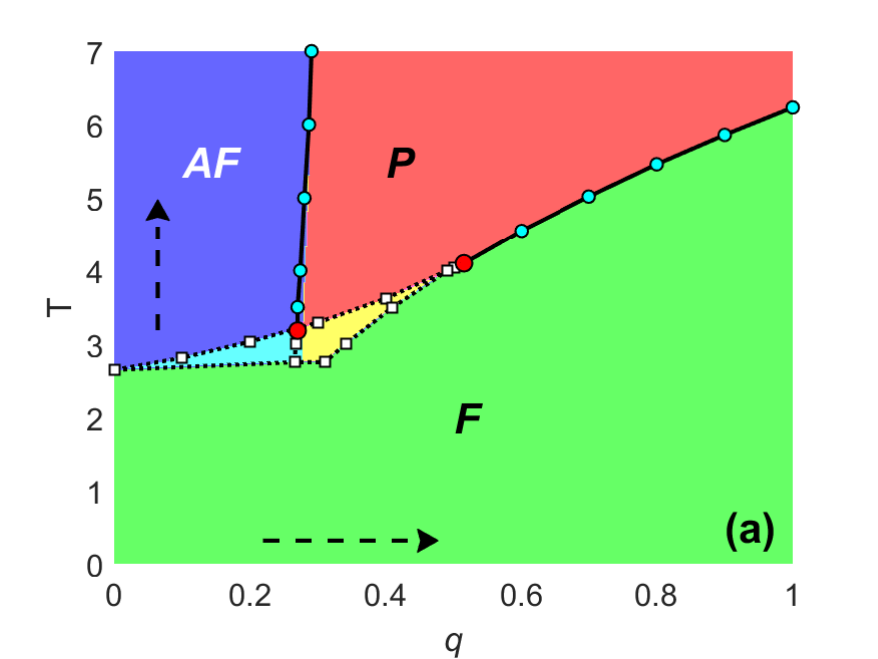}\includegraphics[viewport=40bp 0bp 385bp 300bp,clip,scale=0.6]{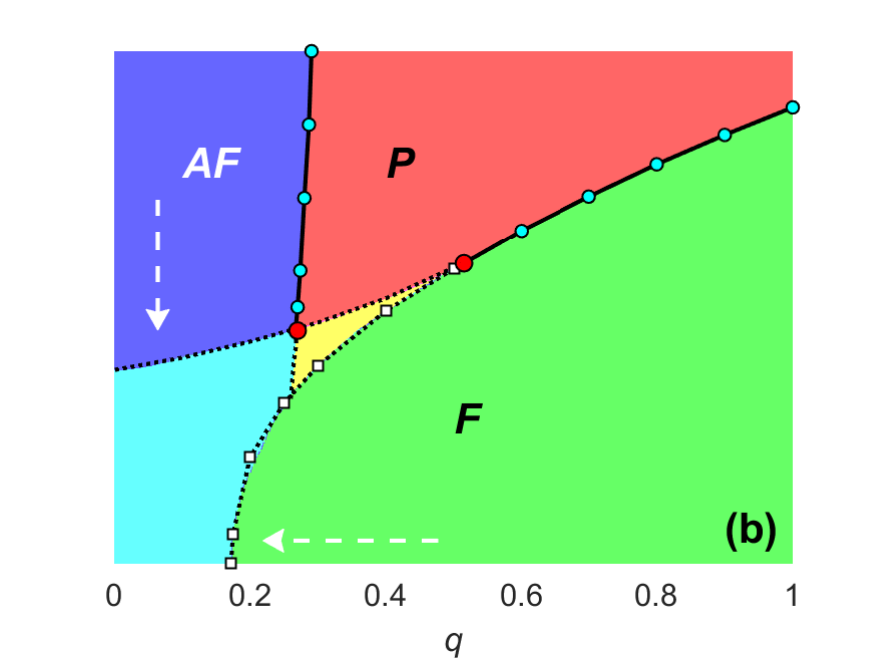}
\par\end{centering}
\caption{{\footnotesize{}Phase diagrams $q$ as a function of $T$ for the
ordered initial state of spins in the simulations. These diagrams
present regions with different colors and denote the phases: $AF$
(purple), $F$ (green), $P$ (red), $AF$ or $F$ (cyan), and $F$
or $P$ (yellow). The cyan circles connected by solid lines indicate
second-order phase transition points and the white squares connected
by dashed lines indicate first-order phase transition points. The
red points represent the tricritical points, and the black and white
dashed arrows indicate the direction of sweeping of the parameters
$T$ and $q$ in the simulations. The error bars are smaller than
the symbol sizes.\label{fig:1}}}
\end{figure*}
\par\end{center}

In this subsection, the phase diagrams of the system were obtained
with the ordered $F$ initial state in the simulations. In the diagram
of Fig. \ref{fig:1}(a), the phase transition points were found by
varying the external parameters, $T$ or $q$, from the lowest to
the highest value (see black arrows in the figure). At high values
of $q$ and varying $T$, we have observed a second-order phase transition
between the $F$ and $P$ phases. Thus, regardless of the initial
state of the system and the starting point of the simulation, we consistently
obtained the same critical point value. For temperatures $T\gtrsim3.18$
and up to $T=6.235$, we have observed the self-organization phenomena
in the system, where we start from the $AF$ phase, at low values
of $q$, and pass to the disordered $P$ phase when reducing the external
energy flow into the system. However, we found another ordered phase,
the $F$ phase, when the prevailing dynamics involve the system being
in contact with a heat bath, at high values of $q$. A characteristic
of the critical points at high temperatures is that they indicate
a second-order phase transition. On the other hand, at low values
of $q$ and $T$, the first-order phase transitions are found. Regarding
these first-order phase transitions, the cyan-colored region in the
diagram indicates that we can have both the $F$ and $AF$ phases
in this region, depending on the starting point of the simulations.
If, we start the simulation with parameter values in the purple region,
we find the $AF$ phase in the cyan region. Conversely, if we start
the simulations with the parameter values in the cyan region, we only
find the $F$ phase. The yellow region of the diagram indicates that
we can have both the $P$ and $F$ phases in this region, depending
on the initial parameter values of the simulations. If, we start the
simulations with parameter values in the purple region, we find the
$P$ phase in the yellow region. Nevertheless, if we initiated the
simulations with parameter values in the cyan region, we find only
the $F$ phase in the yellow region.

The existence of these phases can be explained by the dynamics implemented
in the system. At low values of $q$, the Kawasaki dynamics prevail
in the system, simulating an external energy flow into the system.
In this dynamics, the order parameter is conserved, so if we have
an initial state with all spins up, as is the case of the diagram
in Fig. \ref{fig:1}(a) at $q=0$, the only possible state is the
$F$ state. However, if $q\ne0$ the system is also influenced by
the dynamics simulating the system in contact with a heat bath at
temperature $T$, so at low temperatures, the $F$ phase is expected.
Now, when the temperature increases, for low values of $q$, the Kawasaki
dynamics that prevail in the system organize it into the $AF$ phase,
the phase of higher energy of the system, as expected, because the
dynamics governed by the Metropolis mechanism altered the spin states,
so is possible to find a different state from $F$. Thus, for high
values of $T$ and $q$, when the dynamics simulating a system in
constant contact with a heat bath prevail in the system, the $P$
phase is found. Additionally, characterizing the first-order phase
transitions, the cyan and yellow regions in the diagram indicate the
instability of these states near the critical point and can be further
identified in the results throughout this work.

In the diagram in Fig. \ref{fig:1}(b) was also obtained with the
ordered $F$ initial state in the simulations, but the points in this
diagram were found by varying the values of the external parameters,
$T$ or $q$, from the highest to lowest (see white arrows in the
figure). In this case, the second-order phase transition points are
the same as those in the diagram in Fig. \ref{fig:1}(a), but the
way we varied the external parameters allows us to observe new regions
due to first-order phase transitions. Similar to the diagram in Fig.
\ref{fig:1}(a), the cyan region indicates both the $AF$ and $F$
phases, depending on the parameter values at the beginning of the
simulation. If, we start the simulation with the parameter $q$ value
in the purple region and decrease $T$, we always find the $AF$ phase,
but if we start the simulation with the $T$ value in the cyan region,
we only find the $F$ phase. The yellow region indicates both the
$P$ and $F$ phases, depending on the parameter values at the beginning
of the simulation. For instance, it we start the simulation with the
parameter values in the red region, the yellow region represents the
$P$ phase. However, if we start the simulation with external parameter
values from the yellow region, we only find the $F$ phase. The cyan
and yellow regions were delimited by the points from the diagram in
Fig. \ref{fig:1}(a). The first-order phase transition points present
in the diagram in Fig. \ref{fig:1}(b) were obtained by fixing $T$
and varying $q$, or fixing $q$ and varying $T$ starting from the
purple or red regions.

The different regions of the diagram in Fig. \ref{fig:1}(b), when
compared to Fig. \ref{fig:1}(a), result of the dynamics involved,
as well as the values of the parameters that start the simulation
and how they are changed. At low values of $q$ and high values of
$T$, if $q\ne0$, the influence of the heat bath on the system is
sufficient to change randomly the spin states. Nonetheless, since
the prevailing dynamics in the system force it towards the state of
higher energy, we can still find the ordered state of the $AF$ phase.
Maintaining low values of $q$ and starting from high values of $T$,
when we decrease the temperature in the system, the influence of the
dynamics simulating the heat bath in the system is insufficient to
obtain an $F$ phase, and only the $AF$ phase is observed. However,
at low values of $q$ and starting from $T$ values in the cyan-colored
region, the temperature in the system is not high enough to have disordered
spins, so the Glauber mechanism keeps the system ordered.

\subsection{Disordered initial state\label{subsec:Disordered-initial-state}}
\begin{center}
\begin{figure*}
\begin{centering}
\includegraphics[viewport=20bp 0bp 385bp 300bp,clip,scale=0.6]{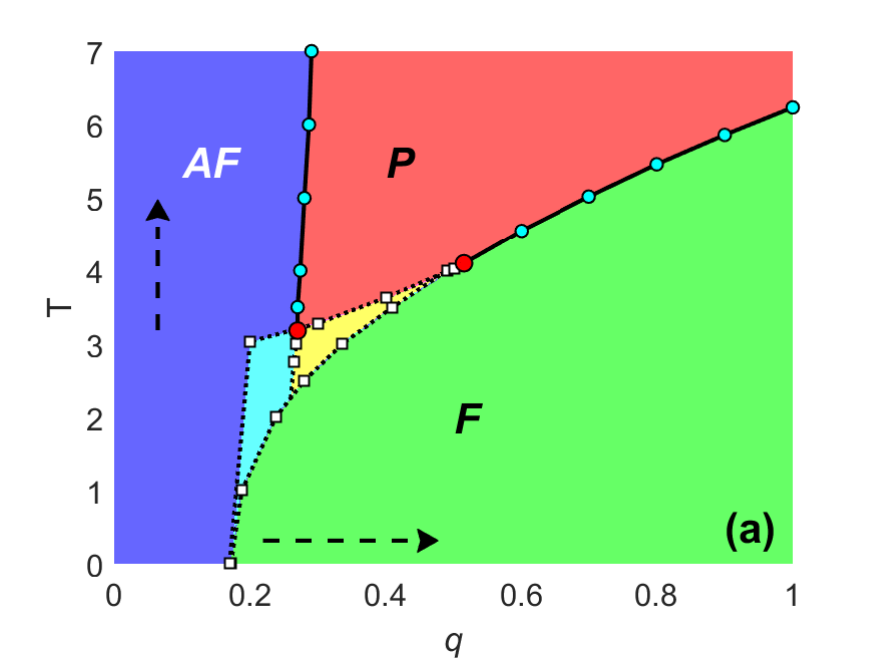}\includegraphics[viewport=40bp 0bp 385bp 300bp,clip,scale=0.6]{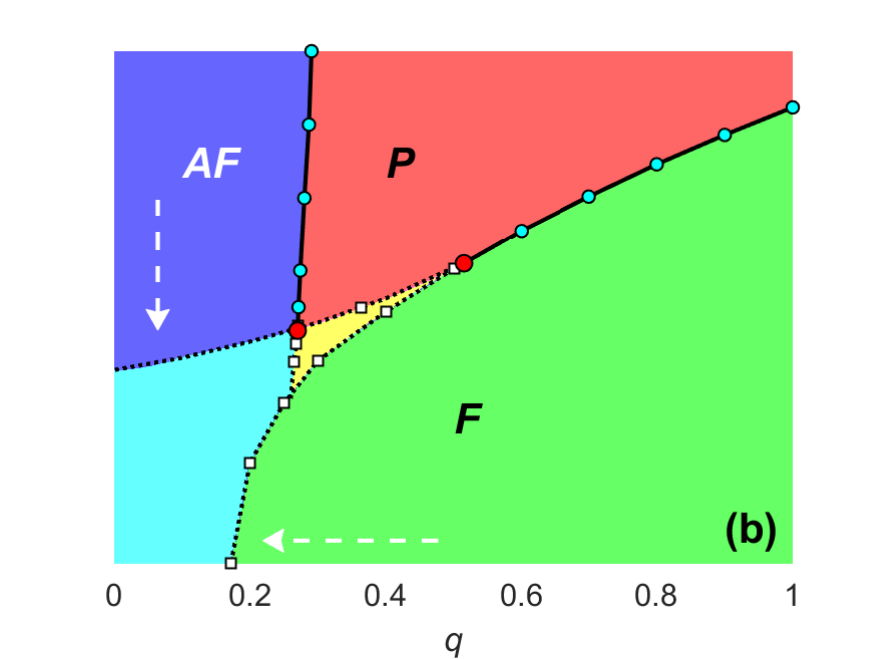}
\par\end{centering}
\caption{{\footnotesize{}Phase diagrams $q$ as a function of $T$ for the
random initial state of spins in the simulations. These diagrams present
regions with different colors and denote the phases: $AF$ (purple),
$F$ (green), $P$ (red), $AF$ or $F$ (cyan), and $F$ or $P$ (yellow).
The cyan circles connected by solid lines indicate second-order phase
transition points and the white squares connected by dashed lines
indicate first-order phase transition points. The red points represent
the tricritical points, and the black and white dashed arrows indicate
the direction of sweeping of the parameters $T$ and $q$ in the simulations.
The error bars are smaller than the symbol sizes. \label{fig:2}}}
\end{figure*}
\par\end{center}

Now, in the phase diagrams of Figs. \ref{fig:2}(a) and (b), we have
the steady states obtained for the system with the random spin initial
state (disordered initial state) in the simulations. In Fig. \ref{fig:2}(a),
the variation of the external parameters, $T$ or $q$, occurs from
the lowest to the highest value (see black arrows in the figure).
When, we have a high external energy flow into the system, i.e., low
values of $q$, only the ordered $AF$ state is found in the system.
Yet, when we increase $q$, the first-order phase transitions from
the $AF$ to $F$ phases are observed. In this diagram, the cyan-colored
region indicates both the $AF$ and $F$ phases, depending on the
value of the external parameter from which we start the simulation.
If, the value of parameter $T$ starts as one of the values in the
green region, the cyan region represents the $F$ phase. However,
if the initial temperature of the system lies between the values of
the cyan region, only the $AF$ phase is found in this region. The
yellow region in the diagram of Fig. \ref{fig:2}(a) can indicate
both $F$ and $P$ phases. Starting the simulation with parameter
values in the green region, the yellow region represents the $F$
phase. Now, if the parameter values are in the cyan or purple region,
the phase found in the yellow region is the $P$ phase. Additionally,
in this diagram the second-order phase transition points are the same
as those found in the diagrams of Figs. \ref{fig:1}(a) and (b).

The cyan and yellow regions in Fig. \ref{fig:2}(a) can be interpreted
as a meta-stable states due to the first-order phase transitions in
this part of the diagram, where we have a greater influence of the
diffusive dynamics, which conserves the order parameter (Kawasaki
dynamics). In this figure, the initial state is one where the spin
states are randomly distributed on the lattice sites. At low values
of $q$, the high energy flow into the system ensures that we always
obtain the state of higher energy, given the Hamiltonian of the Ising
model. However, when the energy flow decreases, increasing $q$, the
dynamics favoring the lower energy state prevail in the competition
between dynamics, and we begin to find the $F$ state in the system
for low values of $T$.

Finally, the diagram in Fig. \ref{fig:2}(b), we also used the disordered
initial state of the spins in the simulation, but now the sweeping
of the external parameters occur from the highest to lowest values
(see white arrows in the figure). In this diagram, we also have a
region where can have both the $AF$ and $F$ phases. The cyan-colored
region, when fixing $q$ and varying $T$, we only obtained the $AF$
phase. However, when fixing $T$ and varying $q$ and starting from
values in the green region, only the $F$ phase is observed. The yellow
region in this diagram also indicates both the $P$ and $F$ phases.
If, the external parameters in the simulation are initialized with
values from the yellow region, we find the $P$ phase in this region.
On the other hand, if the parameter values at the beginning of the
simulation are in the green region of the diagram, the yellow region
refers to $F$ phase.

In Fig. \ref{fig:2}(b), since the initial state of the spins is random
and the parameters are swept from highest to lowest value, at low
values of $q$, even though it is a dynamics that conserves the order
parameter, such as the Kawasaki dynamics prevailing in the system,
we always find the $AF$ phase. Increasing $q$, we find a first-order
phase transition between the $AF$ and $F$ ordered phases. From these
points, at low values of $T$, we only have the presence of the $F$
phase, since the system is simulated in contact with a heat bath at
fixed $T$, where the one-spin-flip dynamics are prominent for high
values of $q$. In this diagram, the meta-stable states of the first-order
phase transitions in the cyan and yellow regions are also evident,
along with the self-organization phenomenon mentioned in the description
of Fig. \ref{fig:1}(a), which can also be seen in all diagrams of
Figs. \ref{fig:1} and \ref{fig:2}.

\subsection{Second-order phase transitions\label{subsec:Second-order-phase-transitions}}

This subsection aims to present the critical behavior observed in
the second-order phase transitions as can be seen in the diagrams
of Fig. \ref{fig:1} and Fig. \ref{fig:2}. The continuous variation
of the order parameter during the transition from an ordered phase
to the higher symmetry disordered phase can be identified by analyzing
the crossing of the Binder cumulant curves at the phase transition
point \citep{24,25}. In our system, the Ising model on the complex
network with competing dynamics exhibits two regions in the phase
diagram, of $T$ versus $q$, with second-order phase transitions.

The first region is observed at low values of $q$ and high of $T$,
in the transition from the $AF$ to $P$ phase, as $q$ increases.
In this part of the diagram, there is a high external energy flow
into the system. Therefore, since the dynamics responsible for this
energy flow favor the state of higher energy in the system, the$AF$
phase is expected to occur. This $AF$ phase is only observed at high
values of $T$ in a second-order phase transition, because, due to
the dynamics simulating the contact with the heat bath in the system,
the only possible phase would be the disordered $P$ phase. Consequently,
it prevents us from having a transition between two ordered phases,
which is one of the main reasons why we find first-order phase transitions
in the system.

Now, the second region where we found second-order phase transitions
in the diagrams of Figs. \ref{fig:1} and \ref{fig:2} is the one
with high values of $q$. In this case, the dominant dynamics in the
system simulate the contact with the heat bath through the one-spin
flip mechanism. This dynamics does not conserve the order parameter.
Therefore, given favorable conditions, i.e., low temperatures, we
will always find the state of lower energy, the $F$ phase, independently
of the initial state of the system. This characteristic of the dynamics
in the system prevents the existence of meta-stable states, and we
can observe continuous phase transitions between the $F$ to $P$
phases.

The Binder cumulant curves for the two types of second-order phase
transitions observed are present in Fig. \ref{fig:3}. Fig. \ref{fig:3}(a),
we display the crossing of the curves indicating the transition point
between $AF$ to $P$ phase as a function of $q$ and for a fixed
value of $T=5.0$. At this same temperature, further increasing $q$,
is observed transitions from the $P$ to $F$ phases, as indicated
by the crossing of the Binder cumulant curves in Fig. \ref{fig:3}(b).
Transitioning from the ordered $AF$ phase to the disordered $P$
phase and from this disordered phase back to an ordered phase ($F$
phase), we can observe the phenomenon of self-organization in our
non-equilibrium system.

Another property that we can obtain from the second-order phase transitions
is the universality class of the system. This universality class can
be identified through the critical exponents, which they were obtained
here through the scale relations in Eqs. (\ref{eq:12}), (\ref{eq:13})
and (\ref{eq:14}). Using these scale relations, there are two main
methods to obtain the exponents of the system.
\begin{center}
\begin{figure}
\begin{centering}
\includegraphics[scale=0.6]{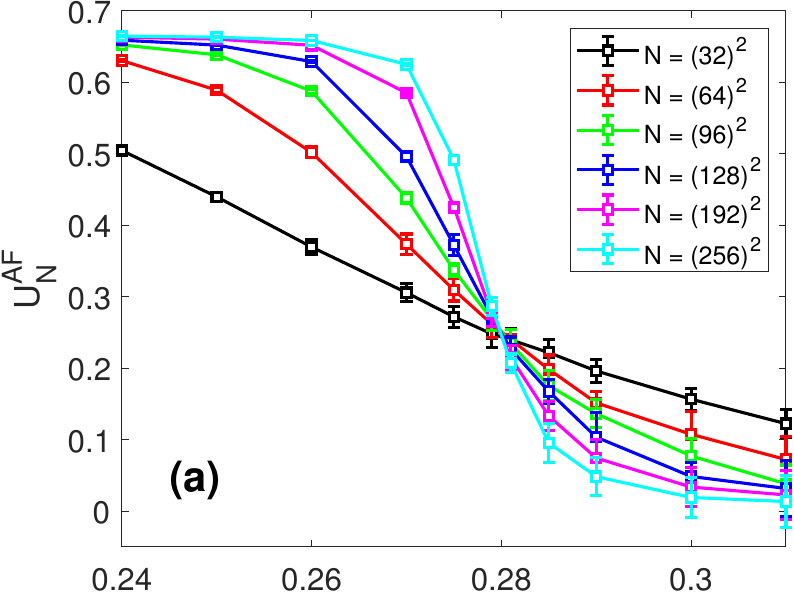}
\par\end{centering}
\begin{centering}
\vspace{0.25cm}
\par\end{centering}
\begin{centering}
\includegraphics[scale=0.6]{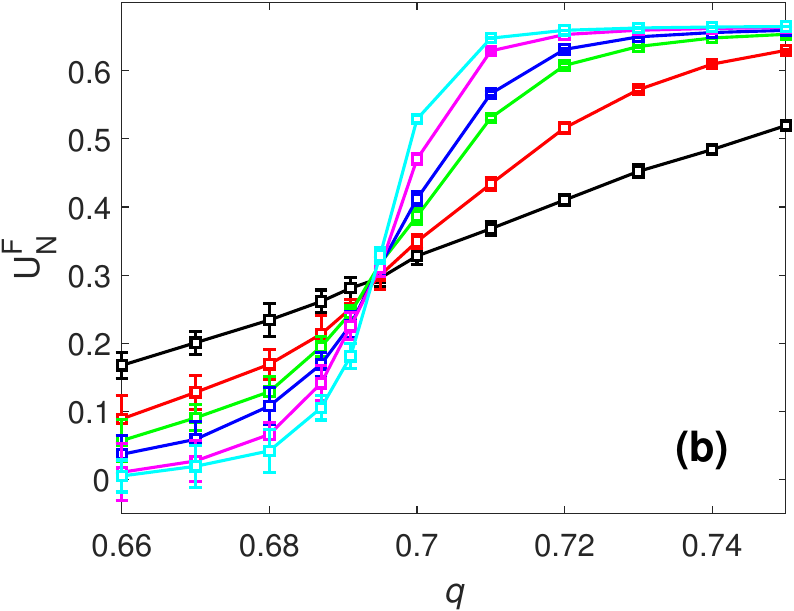}
\par\end{centering}
\caption{{\footnotesize{}Binder cumulants $U_{N}^{AF}$ (a) and $U_{N}^{F}$
(b) as a function of the competition parameter $q$, and for a fixed
value of $T=5.0$. The crossing point for different network sizes
$N$ (see in the figure) indicate the second-order phase transition
point in $q_{c}=0.28\pm0.005$(a) and $q_{c}=0.694\pm0.005$(b). \label{fig:3}}}
\end{figure}
\par\end{center}

The first method can be seen in Fig. \ref{fig:4}, where we have utilized
the fact that scale relations are valid in the vicinity of the critical
point. By collecting data of thermodynamic quantities at the phase
transition for different lattice sizes $N$, the slope of the linear
fit of these points on a graph with axes in logarithmic scale returns
the ratios between the critical exponents. Using scale relation of
Eq. (\ref{eq:12}), the points of the magnetization at $q_{c}$ as
a function of $N$ yield the ratio $\beta/\nu$, as indicated by the
linear fit of the black points in Fig. \ref{fig:4}. Similarly, using
the relation of Eq. (\ref{eq:13}), the susceptibility points show
us the ratio $\gamma/\nu$, while with Eq. (\ref{eq:14}), utilizing
data from the derivative of the Binder cumulant we obtain information
about the exponent related to the correlation length, $1/\nu$. The
linear fit for the ratios $\gamma/\nu$ and $1/\nu$ can be seen respectively
in the red and green points in Fig. \ref{fig:4}. Additionally, in
this figure, the square points indicate the thermodynamic quantities
at the transition between the $AF$ to $P$ phases, while circle points
denote the quantities at the transition between the $P$ to $F$ phases.
In these two transitions, the equivalent critical exponents are obtained,
for $T=5.0$, at the $AF$ to $P$ phase transition we found $\left(\beta/\nu\right)_{AF}=0.24\pm0.01$,
$\left(\gamma/\nu\right)_{AF}=0.51\pm0.02$, and $\left(1/\nu\right)_{AF}=0.51\pm0.09$,
while at the $P$ to $F$ phase transition we have obtained $\left(\beta/\nu\right)_{F}=0.26\pm0.06$,
$\left(\gamma/\nu\right)_{F}=0.50\pm0.06$, and $\left(1/\nu\right)_{F}=0.49\pm0.09$.
\begin{center}
\begin{figure}
\begin{centering}
\includegraphics[scale=0.6]{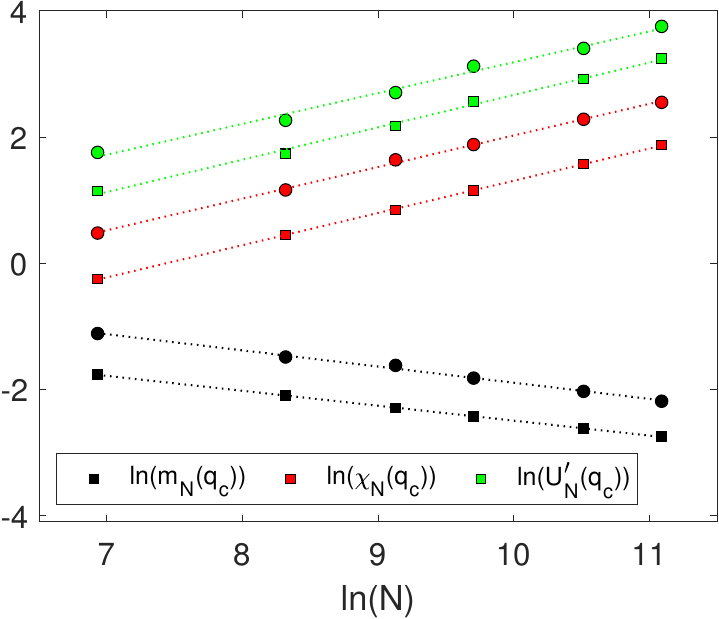}
\par\end{centering}
\caption{{\footnotesize{}Linear fit of the thermodynamic quantities at the
critical point as a function of the network size $N$, and for a fixed
value of $T=5.0$. The square symbols represent the values of the
$m_{N}^{AF}$, $\chi_{N}^{AF}$, and $U_{N}^{AF}$, already the circle
symbols indicate the values of the $m_{N}^{F}$, $\chi_{N}^{F}$,
and $U_{N}^{F}$. The results are well fitted by dashed straight lines.
The error bars are smaller than the symbol sizes.\label{fig:4}}}
\end{figure}
\par\end{center}

Another method that we can employ to find the critical exponents of
the system is data collapse. This method aims to find the scaling
function contained in the scaling relations by collapsing the data
of thermodynamic quantities with different $N$. To achieve this,
isolating the scaling function from the scaling relations, i.e., plotting
$m_{N}N^{\beta/\nu}$ against $\epsilon N^{1/\nu}$ for magnetization
curves, we adjusted the critical exponents until obtaining a single
curve with the different $N$. When this occurs, the exponents used
in this data collapse are the critical exponents of the system, since
the scaling function is only obtained in the vicinity of the critical
point and if the correct critical exponents of the system are used.
In Fig. \ref{fig:5}(a), we present the data collapse of the magnetization,
$m_{N}^{AF}$ and $m_{N}^{F}$, while in Fig. \ref{fig:5}(b), the
data collapse of the susceptibility in the two transitions can be
seen. These plots were made with logarithmic scale axes as this also
allows us to identify the asymptotic behavior, far from $q_{c}$,
of the scaling functions through the slope $\Theta$ presented in
Figs. \ref{fig:5}(a) and (b). Fixed at $T=5.0$, for the data collapse
at the $P-AF$ transition, we used $\left(\beta/\nu\right)_{AF}=0.25$,
$\left(\gamma/\nu\right)_{AF}=0.50$, $\left(1/\nu\right)_{AF}=0.50$,
e $q_{c}=0.28\pm0.005$, and at the $F-P$ transition, we have used
the exponents $\left(\beta/\nu\right)_{F}=0.23$, $\left(\gamma/\nu\right)_{F}=0.52$,
e $\left(1/\nu\right)_{F}=0.50$, along with the phase transition
point $q_{c}=0.694\pm0.005$.
\begin{center}
\begin{figure}
\begin{centering}
\includegraphics[scale=0.6]{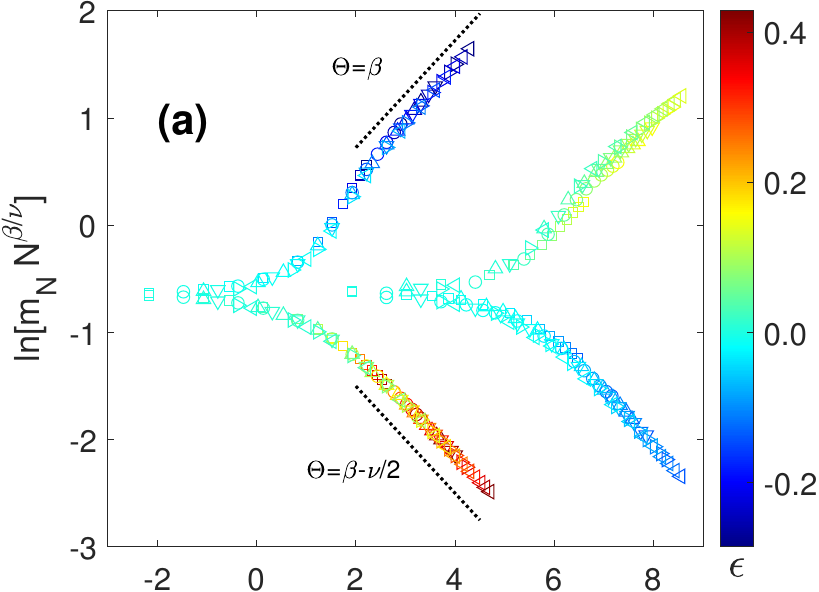}
\par\end{centering}
\begin{centering}
\vspace{0.25cm}
\par\end{centering}
\begin{centering}
\includegraphics[scale=0.6]{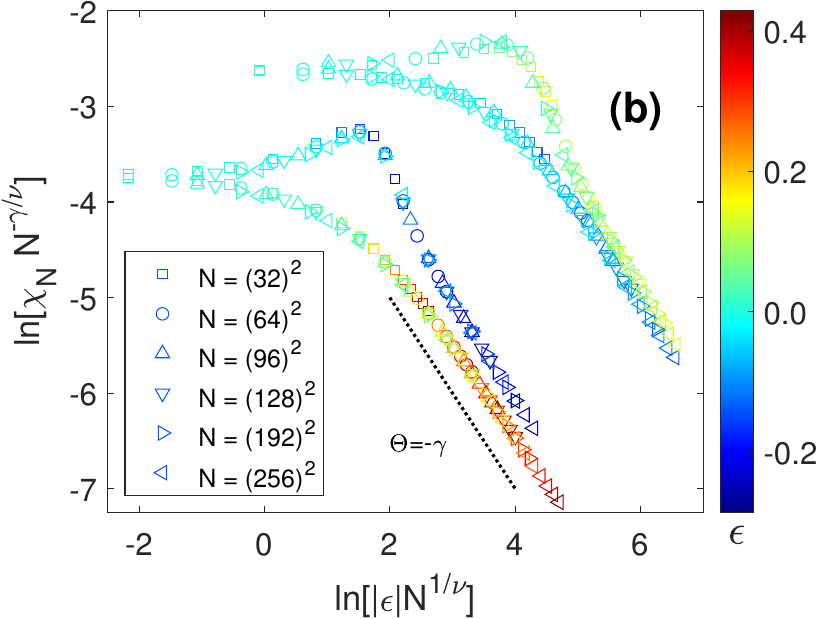}
\par\end{centering}
\caption{{\footnotesize{}Data collapse of the magnetizations (a) and susceptibilities
(b) for different network sizes $N$, as indicated in the figure,
and for a fixed value of $T=5.0$. The curves of $m_{N}^{AF}$ and
$\chi_{N}^{AF}$ can be seen in the right side of the figures, and
for $m_{N}^{F}$ and $\chi_{N}^{F}$ the collapsed curves are in the
left side of the figures. The data collapse validates our estimates
for the critical parameters $\beta/\nu$ , $\gamma/\nu$, $1/\nu$,
and$q_{c}$. The error bars are smaller than the symbol sizes.\label{fig:5}}}
\end{figure}
\par\end{center}

In both methods used for calculating the critical exponents, we obtained
equivalent results, and we can say that the system at the second-order
phase transition belongs to the universality class of mean-field approximation,
since $\beta=0.5$, $\gamma=1.0$, and $\nu=2.0$. These exponents
are expected because we are dealing with the Ising model on a complex
network where the second and fourth moments of the degree distribution
are convergent. With this result, we have further evidence that the
Ising model belongs to the same universality class both in thermodynamic
equilibrium and out of it \citep{21}.

\subsection{First-order phase transitions\label{subsec:First-order-phase-transitions}}

As mentioned in the description of the phase diagrams in Figs. \ref{fig:1}
and \ref{fig:2}, we found first-order transitions at low values of
$q$. In this part of the diagrams, when we decrease $q$, we increase
the external energy flow into the system. In this case, the Kawasaki
dynamics tend to govern the system. As a dynamics that conserves the
order parameter, it depends on initial conditions or complementary
dynamics to reach specific steady states. If, we only have the Kawasaki
dynamics acting on the system $(q=0$), as is predefined to favor
the higher-energy state, and if the initial state of the system is
$F$, then it will not be altered because the dynamics do not change
the spin states. Additionally, if the initial state of the system
is $P$, the system evolves to the $AF$ state, whereas this is the
higher-energy state and the initial spin states do not need to be
altered, only exchanged with each other. Now, if the system is in
an initial $AF$ state, it remains unchanged. On the other hand, if
there is a non-zero probability of the one-spin flip dynamics acting
on the system ($q\ne0$), as is strongly dependent on temperature
and favors the lower-energy state of the system, at high values of
$T$, the spin states are altered to be the disordered state $P$
as the steady state. Yet, if $q$ remains small, at these temperature
values, the state of the system can be organized into the $AF$ phase
because the Kawasaki dynamics still dominates the system. This phase
is observed in all diagrams of Figs. \ref{fig:1} and \ref{fig:2}
at second-order phase transitions.
\begin{center}
\begin{figure}
\begin{centering}
\includegraphics[scale=0.6]{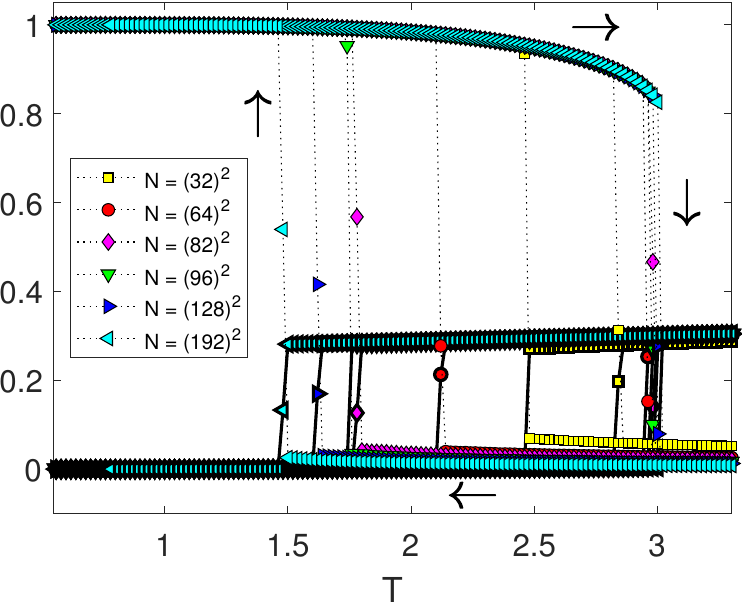}
\par\end{centering}
\caption{{\footnotesize{}Behavior of the magnetization $m_{N}^{F}$ (dotted
lines) and $m_{N}^{AF}$ (solid bold lines) as a function of $T$,
and for different network sizes $N$, as indicated in the figure.
Here, we used the fixed value of $q=0.2$. These hysteresis-like curves
were obtained with the ordered $F$ initial state of the system, we
start from the smallest to the largest value of $T$, and then in
the opposite direction, as indicated by the arrows in the figure.
The error bars are smaller than the symbol sizes. \label{fig:6}}}
\end{figure}
\par\end{center}

For low values of $T$ and $q\ne0$, we have two ordered state possibles
for the system. This is because, from the Glauber dynamics, the steady
state is the $F$ phase, while from the Kawasaki dynamics, we expect
the $AF$ phase. In this case, the initial state of the system makes
a total difference, because, starting from the initial $F$ state
(see Fig. \ref{fig:1} (a)), even with the Kawasaki dynamics as the
most dominant in the system, we do not find any other phase than $F$,
which prevails for all values of $q$. On the other hand, if we have
a disordered initial state (see Fig. \ref{fig:2}(a) and (b)) or we
start from an ordered state, but the system is disordered by the Glauber
dynamics due to high temperatures (see Fig. \ref{fig:1}(b)). When
the Kawasaki dynamics is the most dominant, the system evolves into
the ordered $AF$ state. However, at low temperatures, we have internal
competition in the system between the dynamics, because, this is also
the regime of the $F$ phase in the Glauber dynamics. So, if we decrease
the external energy flow and the system is dominated by the heat bath,
i.e., increasing $q$, an abrupt transition from the $AF$ to $F$
phases is observed. In these transitions between the ordered phases,
we find the majority of the first-order transitions in the system.
The transition between ordered phases can also be observed by changing
the temperature of the system, i.e., at low values of $q$ and initial
$F$ state, increasing $T$, the transition to $AF$ phase is observed.
\begin{center}
\begin{figure}
\begin{centering}
\includegraphics[scale=0.6]{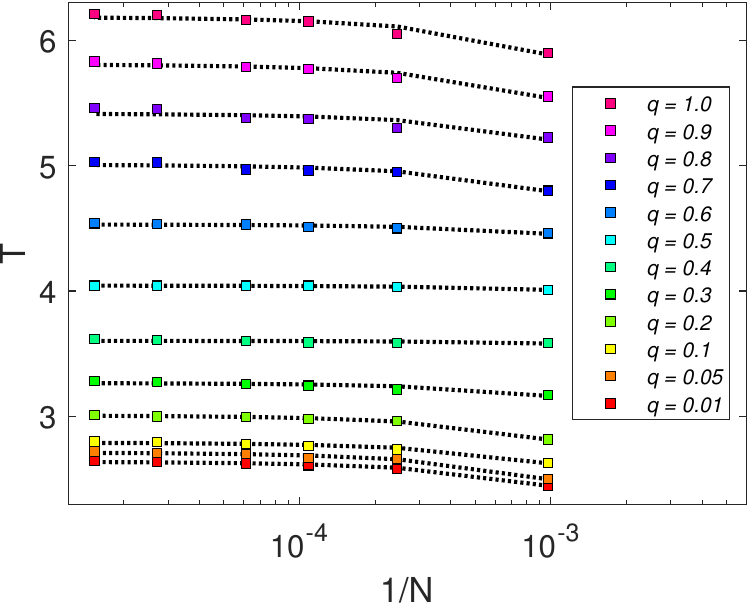}
\par\end{centering}
\caption{{\footnotesize{}Linear adjustment of the temperature where we have
the susceptibility peak, as a function of the inverse of the network
size $N$, to estimate the critical point in the system. These adjustments
were used to calculate the points in Fig. \ref{fig:1} (a), and here
the x-axis is on a logarithmic scale just to better visualize the
points.The error bars are smaller than the symbol sizes. \label{fig:7}}}
\end{figure}
\par\end{center}

Due to the meta-stable states close to first-order phase transitions,
one way to identify these transitions is by examining the dependence
of the transition point on the system size. Another result of this
instability is the possibility of obtaining hysteresis-like curves
by changing the direction of parameter sweeping in first-order phase
transitions. One of the most interesting points to observe these instabilities
is $q=0.2$, because, it passes through all ordered phases and regions
where both the ordered $F$ phase and the ordered $AF$ phase can
exist. In Fig. \ref{fig:6}, for the ordered $F$ initial state in
the system, we present the plot of $m_{N}^{AF}$ and $m_{N}^{F}$,
at $q=0.2$, as a function of $T$, varying from the lowest to the
highest temperature, and then reversing, varying from the highest
to the lowest temperature. Comparing these magnetizations with the
phases obtained in the diagrams of Fig. \ref{fig:1}, we can see that
the approximate point where $m_{N}^{F}$ tends to zero is precisely
on the transition line from the $F$ to $AF$ phase observed in the
diagram of Fig. \ref{fig:1}(a). As we decrease the temperature, we
already have a nonzero value for $m_{N}^{AF}$, transitioning from
the $AF$ to $F$ phase in the vicinity of the transition point in
the diagram of Fig. \ref{fig:1}(b).
\begin{center}
\begin{figure}
\begin{centering}
\includegraphics[scale=0.6]{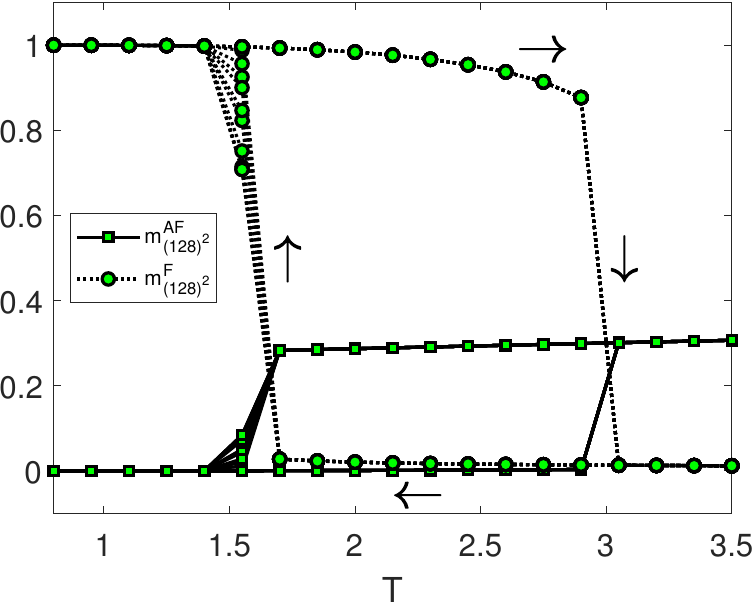}
\par\end{centering}
\caption{{\footnotesize{}Behavior of the magnetizations }$m_{N}^{F}${\footnotesize{}
and }$m_{N}^{AF}${\footnotesize{}as a function of $T$ for network
size $N=(128)^{2}$, and $q=0.2$. We performed $10$ cycles of variation
of $T$, where the initial state of the system was with all spins
in the same state. The error bars are smaller than the symbol sizes.
\label{fig:8}}}
\end{figure}
\par\end{center}

Owing to system size dependence, we can not use the crossing of the
fourth-order Binder cumulant curves to obtain the first-order transition
points of the system. However, we can use the linear behavior of the
peaks of the magnetic susceptibility for different lattice sizes.
In this case, we extrapolated the value of the critical point to an
infinite-sized lattice, assuming that $T_{c}(\chi_{N}^{max})=N^{-1}\Delta+T_{c}(\chi_{N\to\infty}^{max})$,
where $T_{c}(\chi_{N}^{max})$ is the pseudo-critical point for each
network size and $T_{c}(\chi_{N\to\infty}^{max})$ is the extrapolation
of the critical point for infinite network. Thus, we can fit $T_{c}(\chi_{N}^{max})$
as a function of $N^{-1}$ to obtain an estimate for $T_{c}(\chi_{N\to\infty}^{max})$
with the linear coefficient of this fit. Examples of these fits for
different values of $q$ and for both first- and second-order phase
transitions can be seen in Fig. \ref{fig:7}, where we have placed
the $N^{-1}$ axis on a logarithmic scale for better visualization
of the points for the different network sizes after estimating the
phase transition points.

As seen in Fig. \ref{fig:6}, the hysteresis-like curves become even
more unstable when dealing with regions in the phase diagram where
two types of phases can coexist. Another way to observe this characteristic
is by performing multiple loops of the external parameter, in this
case $T$, for $q=0.2$, thus creating several hysteresis-like curves,
as shown in Fig. \ref{fig:8}. In this case, we have a single way
when increasing $T$, but when we reverse the way, decreasing $T$,
in this region where both $F$ and $AF$ phases can exist, we observe
variations regarding the point where the transition between the ordered
phases occurs.

When we decrease the external energy flux into the system, we reduce
the main reason for the existence of first-order phase transitions,
i.e., the coexistence between two ordered phases, as the prevailing
dynamics in the system depends only on temperature and not on the
initial state of the system or auxiliary dynamics. This reduction
in the instability of the system until reaches the second-order phase
transition when $q$ increases can be seen in Fig. \ref{fig:9}: for
the ferromagnetic magnetization curves in Fig. \ref{fig:9}(a), Binder
cumulant in Fig. \ref{fig:9}(b), and magnetic susceptibility in Fig.
\ref{fig:9}(c). In this figure, the hysteresis-like curves were constructed
somewhat differently from those in Fig. \ref{fig:6} and \ref{fig:8}.
The curves with square points represent results of the system with
an ordered $F$ initial state, where all sites have the same spin
value, while the points of the curves with circles represent a system
with a random initial state (disordered sate). Additionally, the dashed
curves indicate that $T$ was swept from higher to lower values, while
the solid curves indicate that the temperature was swept from lower
to higher values.

Another interesting point, that we can observe in Fig. \ref{fig:9}
and in the phase diagrams of Figs. \ref{fig:1} and \ref{fig:2},
is the existence of absorbing states which can be found at $q\lesssim0.172$.
In the case of $q=0.1$ in Fig. \ref{fig:9}, we have an example of
an absorbing state, i.e., starting from the ordered state and increasing
the temperature, looking at $m_{N}^{F}$, we have a transition from
$F$ to $P$ phase, but when we return in the opposite direction by
decreasing the temperature, we have not reached the $F$ phase again.
However, if we start from a disordered initial state, characteristic
of $P$ phase, we never reach the $F$ phase. Another absorbing state,
that can be identified at $q\lesssim0.172$, is not shown here, but
can be analyzed in Figs. \ref{fig:1} and \ref{fig:2}. Looking at
$m_{N}^{AF}$, if initially all spins in the system are in the same
state, increasing the temperature transitions from $P$ to $AF$ phase
is observed. Now, if we reverse the way by decreasing the temperature,
we do not reach the $P$ phase again, only observing $AF$ phase.
Starting from a disordered initial state and still considering $m_{N}^{AF}$,
we always found phase $AF$ as the steady-state. This indicates that
for $q\lesssim0.172$, looking at $m_{N}^{F}$, once the $P$ phase
is reached, we can not leave this phase, while looking at $m_{N}^{AF}$,
once the $AF$ phase is reached, we can not leave this phase either.
This characterizes the absorbing states and can be further verified
in the phase diagrams of Figs. \ref{fig:1} and \ref{fig:2} for $q\lesssim0.172$.
\begin{center}
\begin{figure}
\begin{centering}
\includegraphics[scale=0.6]{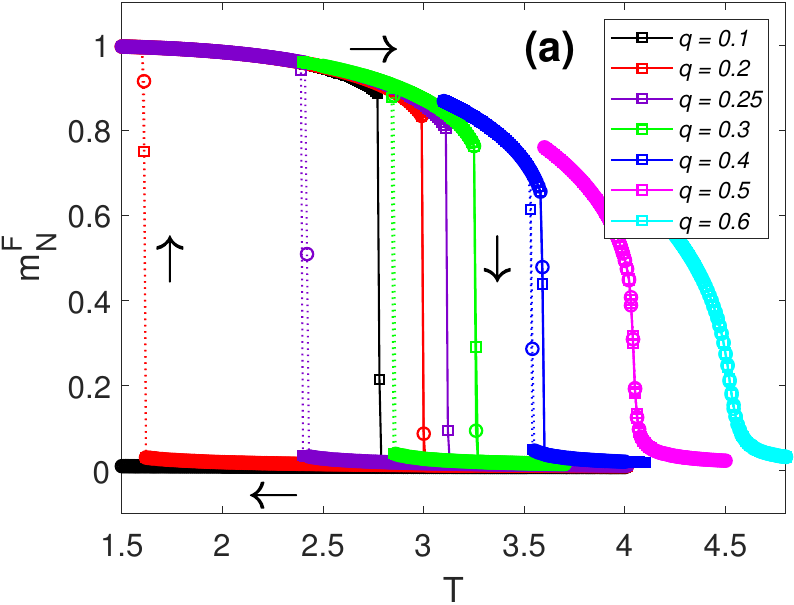}
\par\end{centering}
\begin{centering}
\vspace{0.25cm}
\par\end{centering}
\begin{centering}
\includegraphics[scale=0.6]{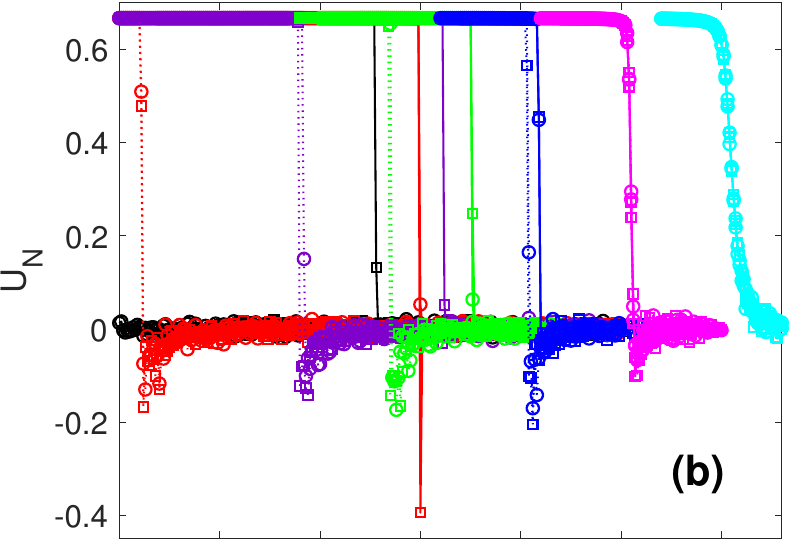}
\par\end{centering}
\begin{centering}
\vspace{0.25cm}
\par\end{centering}
\begin{centering}
\includegraphics[scale=0.6]{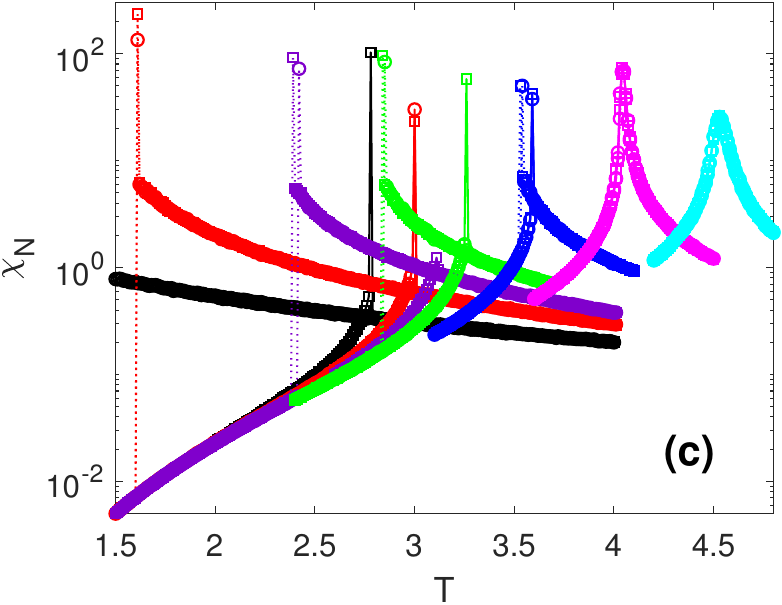}
\par\end{centering}
\caption{{\footnotesize{}Behavior of the thermodynamic quantities of the system
$m_{N}^{F}$ (a), $U_{N}$ (b), and $\chi_{N}$ (c) as a function
of $T$. Here is presented curves for different values of $q$, as
indicated in the figure, and for network size $N=(128)^{2}$. The
solid curves denote the sweep of $T$ from the smallest to the largest
value, while the dotted curves denote the sweep of $T$ from the largest
to the smallest value. The curves with square points indicate that
the system had an ordered $F$ initial state, while in the curves
with circular points the initial state of the system is the disordered
sate. The error bars are smaller than the symbol sizes. \label{fig:9}}}
\end{figure}
\par\end{center}

In the phase diagrams of Figs. \ref{fig:1} and \ref{fig:2}, we have
found both first- and second-order phase transitions and the point
where one type of transition starts and the other ends is where we
can identify as the tricritical point. We do not have a very precise
technique to define the tricritical point, but we can analyze some
evidence that characterizes these types of phase transitions (first-
and second-order) and estimate the value of this point. The most common
characteristic between these two types of phase transitions is the
continuity of the order parameter. For low values of $q$, we have
a discontinuity of the order parameter in the first-order phase transition,
while the continuous phase transition for high values of $q$ characterizes
the second-order phase transitions (see Fig. \ref{fig:9}). In addition
to the discontinuity in first-order phase transitions, we also have
evidence of coexistence between the ordered and the disordered phases,
which can be verified with the distribution of the order parameter
$\rho\left(m_{N}^{F}\right)$ in the vicinity of the critical point.
Therefore, from Monte Carlo simulations, even if we apparently can
not observe the discontinuity of the order parameter, when we analyze
$\rho\left(m_{N}^{F}\right)$ in the vicinity of the critical point
and observe the coexistence of phases and we can identify this as
a first-order phase transition.

In Fig. \ref{fig:10}(a), we display the $\rho\left(m_{N}^{F}\right)$
for some values of $q$ in the vicinity of the transition point. We
can see $\rho\left(m_{N}^{F}\right)$ on the left side before $T_{c}$
and $\rho\left(m_{N}^{F}\right)$ on the right side after $T_{c}$.
Before $T_{c}$, we can observe two maximum points representing the
symmetric values of the order parameter, $\pm m_{N}^{F}$. After $T_{c}$,
we have only one maximum point representing only the disordered phase,
where $m_{N}^{F}=0$. For low values of $q$ in Fig. \ref{fig:10}(a)
and at the phase transition, we can observe three maximum points,
indicating the coexistence of ordered and disordered phases, characteristic
of a first-order phase transition. When we increase the values of
$q$, we can not distinguish the peaks corresponding to the coexistence
phases, indicating that we have a second-order phase transition. For
instance, based on these distributions, we have estimated the tricritical
point of the system given by $q_{t}=0.515\pm0.01$ for $T=4.10$.
Another tricritical point present in the phase diagrams of Figs. \ref{fig:1}
and \ref{fig:2} is related to the transitions from the $AF$ to $F$
phase, and it was also estimated by this method, where we obtained
$T_{t}=3.18\pm0.02$ for $q=0.27$.
\begin{center}
\begin{figure}
\begin{centering}
\includegraphics[scale=0.6]{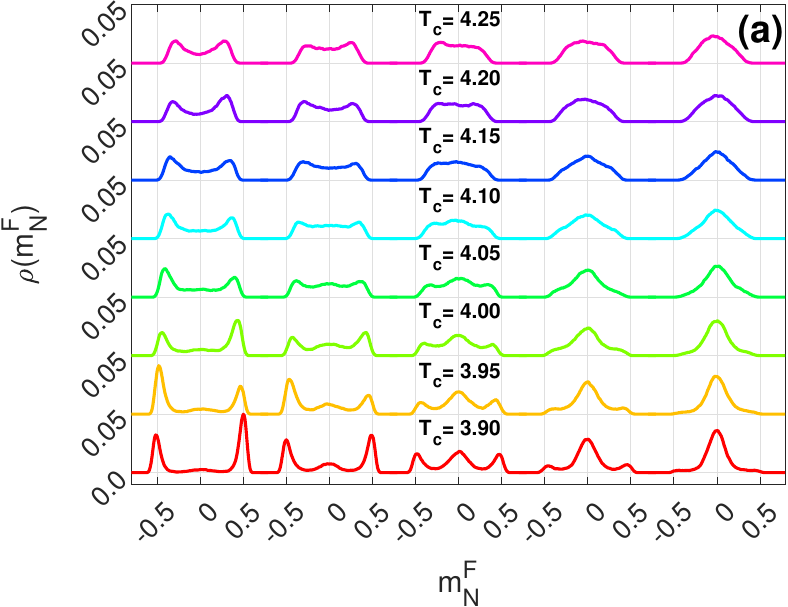}
\par\end{centering}
\begin{centering}
\vspace{0.25cm}
\par\end{centering}
\begin{centering}
\includegraphics[scale=0.6]{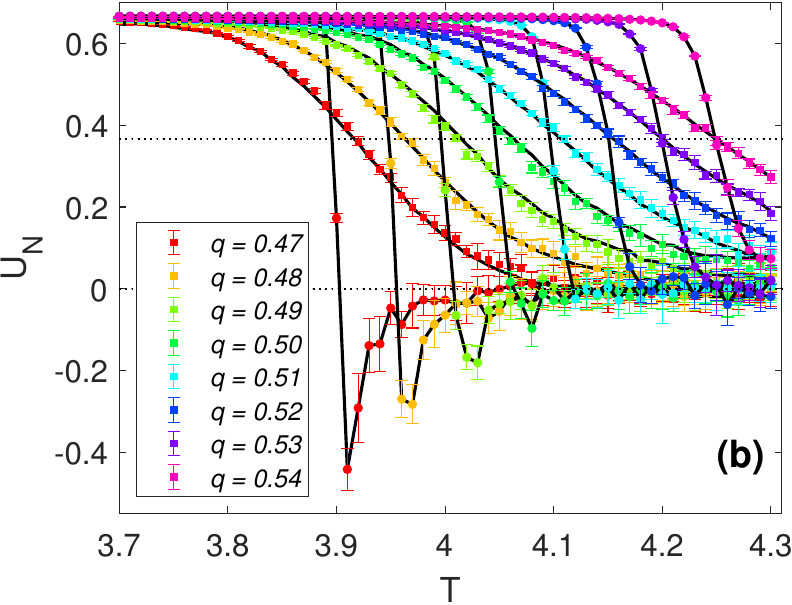}
\par\end{centering}
\caption{{\footnotesize{}(a) The distribution $\rho\left(m_{N}^{F}\right)$
is presented for five values of $T$ around $T_{c}$ and values of
$q$ near the tricritical point. Here, we used $N=(128)^{2}$. (b)
Binder cumulant $U_{N}$ for $m_{N}^{F}$ with two network sizes $L=(32)^{2}$
(square symbols) and $L=(128)^{2}$ (circle symbols). The horizontal
dashed lines are for $U_{N}=0$ and the crossing of the curves for
$q=0.54$ at $U_{N}=0.367$. \label{fig:10}}}
\end{figure}
\par\end{center}

We can also make an observation regarding the values of the Binder
cumulant at the first-order phase transition. For this, with the same
values of $q$ as in the distributions of Fig. \ref{fig:10}(a), we
present the curves for two different network sizes for the Binder
cumulant in Fig. \ref{fig:10}(b). With these curves, we can see that
the Ising model on a complex network with competitive dynamics and
in the first-order phase transitions, also presents negative values
of the Binder cumulant. The dashed lines in Fig. \ref{fig:10}(b)
indicate references at $U_{N}=0$, and the point where the curves
cross for $q=0.54$, $U_{N}=0.367$. This last reference line indicates
that the crossing of the $U_{N}$ curves changes their values, and
this remains even in the second-order phase transitions, and when
$q=1.0$ the crossing is at $U_{N}=0.26$.

\section{Conclusions\label{sec:Conclusions}}

We have employed Monte Carlo simulations to investigate the Ising
model on a network with power-law degree distribution, subject to
two competing dynamics. Considering the ferromagnetic coupling between
spins, with probability $q$, the system is governed by Glauber dynamics,
favoring the lowest energy state, while with probability $1-q$, the
Kawasaki dynamics evolves the system towards the highest energy state.
Given that Kawasaki dynamics conserves the order parameter, we built
the phase diagrams $T$ versus $q$ with different initial states
in the simulations. The topology of theses diagrams revealed regions
with both first- and second-order phase transitions, leading to the
discovery of tricritical points at coordinates ($q=0.27$,$T_{t}=3.18\pm0.02$)
and ($q_{t}=0.515\pm0.01$,$T=4.1$). For $3.18\leq T\leq6.235$ the
self-organization phenomena is observed in the system. Here, at low
$q$ values, the system exhibits the $AF$ phase, transitioning to
the $P$ phase as $q$ increases, and further transitioning to an
ordered phase, $F$ phase, for higher $q$. In regions of second-order
phase transitions, we found that the universality class of the system
is of mean-field approximation, with the critical exponents $\nu=2$,
$\gamma=1$ and $\beta=0.5$. This universality class was expected
due to dealing with networks where the second and fourth moments of
the degree distribution are convergent \citep{19,21}. However, in
these phase diagrams, we also observed first-order phase transitions,
not previously observed in systems with dynamics that do not conserve
the order parameter \citep{18,19}, or systems with competing Glauber
and Kawasaki dynamics, but in the regular networks\citep{8}. This
region, characterized by a discontinuity in the order parameter, arises
due to the competition between $AF$ and $F$ ordered phases. This
is because for low $q$ and $T$, the system tends to organize into
a $AF$ phase due to the strong influence of Kawasaki dynamics, but
for this, initially a disordered state of the system is necessary.
On the other hand, simultaneously, as we are in the regime of low
temperatures, even with the limited influence of Glauber dynamics,
it is still possible to find the $F$ phase when starting with ordered
initial state. Lastly, we have identified absorbing states in the
system for below $q\lesssim0.172$. Starting from the $F$ phase,
at high $T$ the system transits to the $P$ phase when observing
$m_{N}^{F}$, but when decreasing the temperature $T$, the $F$ phase
is not recovered. Similarly, when observing $m_{N}^{AF}$, this absorbing
state is the $AF$ phase, where now, analogously, we can not reach
the $P$ phase at low $T$.

\begin{acknowledgments}
This work was financially supported by the Conselho Nacional de Desenvolvimento
Científico e Tecnológico (CNPq) of Brazil.
\end{acknowledgments}

\end{document}